\documentstyle[twoside]{article}

\input epsf.tex

\catcode`\@=11
\long\def\@makefntext#1{
\protect\noindent \hbox to 3.2pt {\hskip-.9pt  
$^{{\eightrm\@thefnmark}}$\hfil}#1\hfill}		%CAN BE USED 

\def\@makefnmark{\hbox to 0pt{$^{\@thefnmark}$\hss}}	%ORIGINAL 
	
\def\ps@myheadings{\let\@mkboth\@gobbletwo
\def\@oddhead{\hbox{}
\rightmark\hfil\eightrm\thepage}   
\def\@oddfoot{}\def\@evenhead{\eightrm\thepage\hfil
\leftmark\hbox{}}\def\@evenfoot{}
\def\sectionmark##1{}\def\subsectionmark##1{}}

\oddsidemargin=\evensidemargin
\newcounter{sectionc}\newcounter{subsectionc}\newcounter{subsubsectionc}
\renewcommand{\section}[1] {\vspace{12pt}\addtocounter{sectionc}{1} 
\setcounter{subsectionc}{0}\setcounter{subsubsectionc}{0}\noindent 
	{\tenbf\thesectionc. #1}\par\vspace{5pt}}
\renewcommand{\subsection}[1] {\vspace{12pt}\addtocounter{subsectionc}{1} 
	\setcounter{subsubsectionc}{0}\noindent 
	{\bf\thesectionc.\thesubsectionc. {\kern1pt \bfit #1}}\par\vspace{5pt}}
\renewcommand{\subsubsection}[1] {\vspace{12pt}\addtocounter{subsubsectionc}{1}
	\noindent{\tenrm\thesectionc.\thesubsectionc.\thesubsubsectionc.
	{\kern1pt \tenit #1}}\par\vspace{5pt}}

\newcounter{appendixc}
\newcounter{subappendixc}[appendixc]
\newcounter{subsubappendixc}[subappendixc]
\renewcommand{\thesubappendixc}{\Alph{appendixc}.\arabic{subappendixc}}
\renewcommand{\thesubsubappendixc}
	{\Alph{appendixc}.\arabic{subappendixc}.\arabic{subsubappendixc}}

\renewcommand{\appendix}[1] {\vspace{12pt}
        \refstepcounter{appendixc}
        \setcounter{figure}{0}
        \setcounter{table}{0}
        \setcounter{lemma}{0}
        \setcounter{theorem}{0}
        \setcounter{corollary}{0}
        \setcounter{definition}{0}
        \setcounter{equation}{0}
        \renewcommand{\thefigure}{\Alph{appendixc}.\arabic{figure}}
        \renewcommand{\thetable}{\Alph{appendixc}.\arabic{table}}
        \renewcommand{\theappendixc}{\Alph{appendixc}}
        \renewcommand{\thelemma}{\Alph{appendixc}.\arabic{lemma}}
        \renewcommand{\thetheorem}{\Alph{appendixc}.\arabic{theorem}}
        \renewcommand{\thedefinition}{\Alph{appendixc}.\arabic{definition}}
        \renewcommand{\thecorollary}{\Alph{appendixc}.\arabic{corollary}}
        \renewcommand{\theequation}{\Alph{appendixc}.\arabic{equation}}
        \noindent{\tenbf Appendix \theappendixc #1}\par\vspace{5pt}}
\newcommand{\subappendix}[1] {\vspace{12pt}
        \refstepcounter{subappendixc}
        \noindent{\bf Appendix \thesubappendixc. {\kern1pt \bfit #1}}
	\par\vspace{5pt}}
\newcommand{\subsubappendix}[1] {\vspace{12pt}
        \refstepcounter{subsubappendixc}
        \noindent{\rm Appendix \thesubsubappendixc. {\kern1pt \tenit #1}}
	\par\vspace{5pt}}

\newcommand{\textlineskip}{\baselineskip=13pt}
\newcommand{\smalllineskip}{\baselineskip=10pt}

\def\eightcirc{
\begin{picture}(0,0)
\put(4.4,1.8){\circle{6.5}}
\end{picture}}
\def\eightcopyright{\eightcirc\kern2.7pt\hbox{\eightrm c}} 

\newcommand{\copyrightheading}[1]
	{\vspace*{-2.5cm}\smalllineskip{\flushleft
	{\footnotesize International Journal of Modern Physics A, #1}\\
	{\footnotesize $\eightcopyright$\, World Scientific Publishing
	 Company}\\
	 }}

\def\abstracts#1#2#3{{
	\centering{\begin{minipage}{4.5in}\baselineskip=10pt\footnotesize
	\parindent=0pt #1\par 
	\parindent=15pt #2\par
	\parindent=15pt #3
	\end{minipage}}\par}} 

\renewenvironment{thebibliography}[1]
	{\frenchspacing
	 \ninerm\baselineskip=11pt
	 \begin{list}{\arabic{enumi}.}
	{\usecounter{enumi}\setlength{\parsep}{0pt}
	 \setlength{\leftmargin 12.7pt}{\rightmargin 0pt} %FOR 1--9 ITEMS
	 \setlength{\itemsep}{0pt} \settowidth
	{\labelwidth}{#1.}\sloppy}}{\end{list}}

\newcounter{itemlistc}
\newcounter{romanlistc}
\newcounter{alphlistc}
\newcounter{arabiclistc}

\newcommand{\fcaption}[1]{
        \refstepcounter{figure}
        \setbox\@tempboxa = \hbox{\footnotesize Fig.~\thefigure. #1}
        \ifdim \wd\@tempboxa > 5in
           {\begin{center}
        \parbox{5in}{\footnotesize\smalllineskip Fig.~\thefigure. #1}
            \end{center}}
        \else
             {\begin{center}
             {\footnotesize Fig.~\thefigure. #1}
              \end{center}}
        \fi}

\newcommand{\tcaption}[1]{
        \refstepcounter{table}
        \setbox\@tempboxa = \hbox{\footnotesize Table~\thetable. #1}
        \ifdim \wd\@tempboxa > 5in
           {\begin{center}
        \parbox{5in}{\footnotesize\smalllineskip Table~\thetable. #1}
            \end{center}}
        \else
             {\begin{center}
             {\footnotesize Table~\thetable. #1}
              \end{center}}
        \fi}

\def\@citex[#1]#2{\if@filesw\immediate\write\@auxout
	{\string\citation{#2}}\fi
\def\@citea{}\@cite{\@for\@citeb:=#2\do
	{\@citea\def\@citea{,}\@ifundefined
	{b@\@citeb}{{\bf ?}\@warning
	{Citation `\@citeb' on page \thepage \space undefined}}
	{\csname b@\@citeb\endcsname}}}{#1}}

\newif\if@cghi
\def\cite{\@cghitrue\@ifnextchar [{\@tempswatrue
	\@citex}{\@tempswafalse\@citex[]}}
\def\citelow{\@cghifalse\@ifnextchar [{\@tempswatrue
	\@citex}{\@tempswafalse\@citex[]}}
\def\@cite#1#2{{$\null^{#1}$\if@tempswa\typeout
	{IJCGA warning: optional citation argument 
	ignored: `#2'} \fi}}

\def\pmb#1{\setbox0=\hbox{#1}
	\kern-.025em\copy0\kern-\wd0
	\kern.05em\copy0\kern-\wd0
	\kern-.025em\raise.0433em\box0}

\def\fnt#1#2{\footnotetext{\kern-.3em
	{$^{\mbox{\scriptsize #1}}$}{#2}}}

\def\fpage#1{\begingroup
\voffset=.3in
\thispagestyle{empty}\begin{table}[b]\centerline{\footnotesize #1}
	\end{table}\endgroup}

\def\runninghead#1#2{\pagestyle{myheadings}
\markboth{{\protect\footnotesize\it{\quad #1}}\hfill}
{\hfill{\protect\footnotesize\it{#2\quad}}}}
\font\tenrm=cmr10
\font\tenit=cmti10 
\font\tenbf=cmbx10
\font\bfit=cmbxti10 at 10pt
\font\ninerm=cmr9

\font\eightrm=cmr8

\def\PLB{{\em Phys. Lett.} B }
\def\PRL{{\em Phys. Rev. Lett. }}
\def\PRD{{\em Phys. Rev.} D }

\def\etal{{\it et.~al.}}

\def\as{\alpha_S}
\def\Ld{\Lambda}
\def\ld{\lambda}
\def\lsim{\stackrel{<}{\scriptstyle \sim}}

\def\be{\begin{equation}}
\def\ee{\end{equation}}
\def\bea{\begin{eqnarray}}
\def\eea{\end{eqnarray}}
\newcommand \bean {\begin{eqnarray*}}
\newcommand \eean {\end{eqnarray*}}
\newcommand \bary {\begin{array}}
\newcommand \eary {\end{array}}
\newcommand \lan {\langle}
\newcommand \ran {\rangle}

\newcommand{\bi}{\bibitem}
\newcommand{\bit}{\begin{itemize}}
\newcommand{\eit}{\end{itemize}}

\def\qed{\hbox{${\vcenter{\vbox{			%HOLLOW SQUARE
   \hrule height 0.4pt\hbox{\vrule width 0.4pt height 6pt
   \kern5pt\vrule width 0.4pt}\hrule height 0.4pt}}}$}}

\begin{document}

\runninghead{Bounds on Heavy-to-Heavy Weak Decay Form Factors
$\ldots$} {Bounds on Heavy-to-Heavy Weak Decay Form Factors $\ldots$}

\normalsize\textlineskip
\thispagestyle{empty}
\setcounter{page}{1}

\copyrightheading{}			%{Vol. 0, No. 0 (1993) 000--000}

\vspace*{0.88truein}

\fpage{1}
\centerline{\bf Bounds on Heavy-to-Heavy Weak Decay Form Factors}
\vspace*{0.37truein}
\centerline{\footnotesize Cheng-Wei Chiang\footnote{
E-mail address: chengwei@andrew.cmu.edu.}}
\vspace*{0.015truein}
\centerline{\footnotesize\it 
Department of Physics, Carnegie Mellon University, 
5000 Forbes Avenue}
\baselineskip=10pt
\centerline{\footnotesize\it 
Pittsburgh, Pennsylvania 15213, USA}
\vspace*{0.225truein}

\vspace*{0.21truein}
\abstracts{We provide upper and lower bounds on the semileptonic weak
decay form factors for $B \to D^(*)$ and $\Lambda_b \to \Lambda_c$
decays by utilizing inclusive heavy quark effective theory sum rules.
These bounds are calculated to second order in $\Lambda_{QCD}/m_Q$ and
first order in $\alpha_s$.  The $O(\alpha_s^2 \beta_0)$ corrections to
the bounds at zero recoil are also presented.}{}{}

\textlineskip			%) USE THIS MEASUREMENT WHEN THERE IS
\vspace*{12pt}			%) NO SECTION HEADING

%\vspace*{1pt}\textlineskip	%) USE THIS MEASUREMENT WHEN THERE IS
%\section{General Appearance}	%) A SECTION HEADING
\vspace*{-0.5pt}

\noindent

Form factors play an important role in both experimental measurements
and theoretical calculations.  In particular, they are often used to
provide theoretical input for extraction of CKM matrix elements such
as $|V_{cb}|$ and $|V_{ub}|$.  However, the form factors used are not
calculated from first principles but taken from models that have some
additional assumptions.  Therefore, it would be desirable if one could
put constraints on the form factors that are free from any model
dependence.  Such bounds using heavy quark effective theory (HQET)
inclusive sum rules had been derived.\cite{B1995,B1997} They
have been further improved and applied to all form factors in $B \to
D^{(*)}$ and $\Ld_b \to \Ld_c$ decays to first order in both
$O(1/m_Q)$ and $O(\as)$ to the full spectrum and to $O(\as^2 \beta_0)$
at zero recoil.\cite{CL1999,C1999} They can provide a test for the
models used in the literatures.

Using the optical theorem for the inclusive decay rate, one finds
\footnote{We follow the notation used by Chiang \cite{C1999} where
one can find a detailed derivation.}, taking the initial state to be a
$B$ meson as an example:
\bea
&& \frac{1}{2\pi i} \int_{C} d\epsilon \, \theta(\Delta-\epsilon) \,
T(\epsilon) \left( 1-\frac{\epsilon}{E_1-E_H} \right) \nonumber
\\
\label{bounds}
&& \leq \frac{\left| \lan H(v') \right| a \cdot J \left| B(v) \ran
   \right|^2}{4M_B E_H}
\leq \frac{1}{2\pi i} \int_{C} d\epsilon \, \theta(\Delta-\epsilon) \, 
   T(\epsilon)
\left( 1-\frac{\epsilon}{E_{max}-E_H} \right).
\eea
One can readily obtain the corresponding formula for baryons by
averaging over spins where appropriate.  The middle part of
Eq.~(\ref{bounds}) contains a hadronic matrix element that involves
long distance physics.  The goal is to find the proper 4-vector
$a_{\mu}$ and weak decay current $J^{\mu}$ to project out the form
factor combination that is of interest.  On both sides of
Eq.~(\ref{bounds}), one performs calculations in the partonic picture.
The moments of $T(\epsilon)$ multiplied by the weight function
$\theta(\Delta-\epsilon)$ can be computed perturbatively in QCD when
the integration contour $C$ is far from the cuts of physical
processes.  One also performs an operator product expansion (OPE) in
powers of the inverse heavy quark mass $1/m_b$ for $T(\epsilon)$.  In
general, if one performs the OPE for $T(\epsilon)$ to $O(1/m_b^n)$,
then, due to consistency, both bounds are correct to
$O(1/m^{n-1})$.\cite{C1999}  The above-derived bounds have the features
that: (i) the upper bound is model independent while the lower bound
assumes that there is little contribution from multi-particle
production, such as $B \to D\,\pi\,{\it l}\,\nu$, which is supported
by experiments; (ii) the bounds can be applied to the whole kinematic
regime in the case of heavy-to-heavy decays.

Schematically, our partonic calculations of the structure functions,
$T_i$, can be expressed as:
\bea
\label{partonic}
T_i^{Full} &\simeq& T_i^1 + T_i^{1/m_Q} + T_i^{1/m_Q^2} + T_i^{1/m_Q^3}
\\ \nonumber
&&
+ \; \alpha_S \left[\,U_i + (\omega-1) V_i\,\right]
+ \; \alpha_S^2\beta_0 \; T_i^{\alpha_S^2\beta_0}(\omega=1),
\eea
where the first line of Eq.~(\ref{partonic}) is an expansion in powers
of $1/m_Q$.\cite{MW1994} We only expand the first order perturbation
to terms linear in $\omega-1$ because it is a good approximation
within the allowed kinematic regime for heavy-to-heavy
decays.\cite{B1997,CL1999} The results for
$T_i^{\alpha_S^2\beta_0}(\omega=1)$ evaluated at zero recoil are
helpful in understanding the convergence of the perturbation series
and are obtained using the method proposed by Smith and
Voloshin.\cite{SV1994} \footnote{To utilize the method, one has to use
a finite gluon mass to regularize the IR divergence} Corrections of
order $(\frac{\Lambda_{\rm QCD}}{m_Q})^3$, $\alpha_S^2$,
$\alpha_S\,\frac{\Lambda_{\rm QCD}}{m_Q}$, and
$\alpha_S\,(\omega-1)^2$ are neglected.  Table 1 lists the expansion
parameters used in the calculations:

\begin{table}[htbp]
\tcaption{Expansion Parameters.}
\centerline{\footnotesize HQET parameters}
\vspace{2pt}
\centerline{\footnotesize
\begin{tabular}{cl}
\hline
$\Lambda_{\rm QCD}$ & $\sim 0.5\;{\rm GeV}$ \\
${\bar \Lambda}/M_B$ &
$\sim 0.1$ for mesons; \cite{G1996}
$\sim 0.15$ for baryons \\
$\lambda_1$ &
$-0.19\pm0.10\;{\rm GeV^2}$ for mesons; \cite{G1996}
$-0.43\pm0.10\;{\rm GeV^2}$ for baryons \\
$\lambda_2$ &
$0.12\;{\rm GeV^2}$ for mesons;
$0$ for baryons \\
$\rho_1$, ${\cal T}_1$ and ${\cal T}_2$ &
$\sim \Lambda_{\rm QCD}^3$ for
mesons and baryons \\
$\rho_2$, ${\cal T}_3$ and ${\cal T}_4$ &
$\sim \Lambda_{\rm QCD}^3$ for mesons;
$0$ for baryons \\
\hline
\end{tabular}}

\vspace{0.06in}

\centerline{\footnotesize Perturbative parameters}
\vspace{2pt}
\centerline{\footnotesize
\begin{tabular}{cc}  
\hline
$m_b$ & $4.8\;{\rm GeV}$ \\
$m_c$ & $1.4\;{\rm GeV}$ \\
$\alpha_S(2\;{\rm GeV})$ & $\sim 0.3$ \\
$\Delta$ & $1\;{\rm GeV}$ \\
\hline
\end{tabular}}
%\vspace{6pt}
\end{table}

As an example, the bounds on $h_{A_1}$ are of particular interest
because the form factor $F(\omega)$ that appears in $B \to D^* {\it l}
\nu$ decay approximates $h_{A_1}$ when $\omega \to 1$.  Information on
this can help determine the CKM matrix element
$|V_{cb}|$.\cite{VS1987} Here one can choose $a^{\mu}=(0,1,0,0)$ and
an axial current $A_{\mu}$ in Eq.~(\ref{bounds}) to form the bounds on
$f(\omega) \equiv (\omega+1)^2 |h_{A_1}|^2/4\omega$.  Full analyses of
other form factors and comparison with models often used or quoted in
the literature are given elsewhere.\cite{BBC2000}

At zero recoil, we find that to first order in $1/m_b$ both the upper
and lower bounds on $f(1)$ coincide at 1, agreeing with Luke's
theorem.\cite{L1990} When $O(\as)$ corrections are included, the
bounds become $0.916 \leq f(1) \leq 0.927$ using the parameters given
in Table 1.  When uncertainties in the HQET parameters are taken into
account, {\it i.e.~} by varying the parameters ${\bar \Lambda} \in
(0.3,0.5)\; {\rm GeV}$, $\ld_1 \in (-0.1,-0.3)\; {\rm GeV^2}$,
$\rho_{1,2}={\cal T}_{1,2,3,4} \in (-0.125,0.125)\; {\rm GeV^3}$ and
keeping $\ld_2=0.12\; {\rm GeV^2}$, the bounds get widened, as shown
in FIG. 1.
\begin{figure}[t]
%\centerline{\vbox{\hrule width 5cm height0.001pt}}
\begin{center}
\epsfysize=4truecm
\centerline{\epsfbox{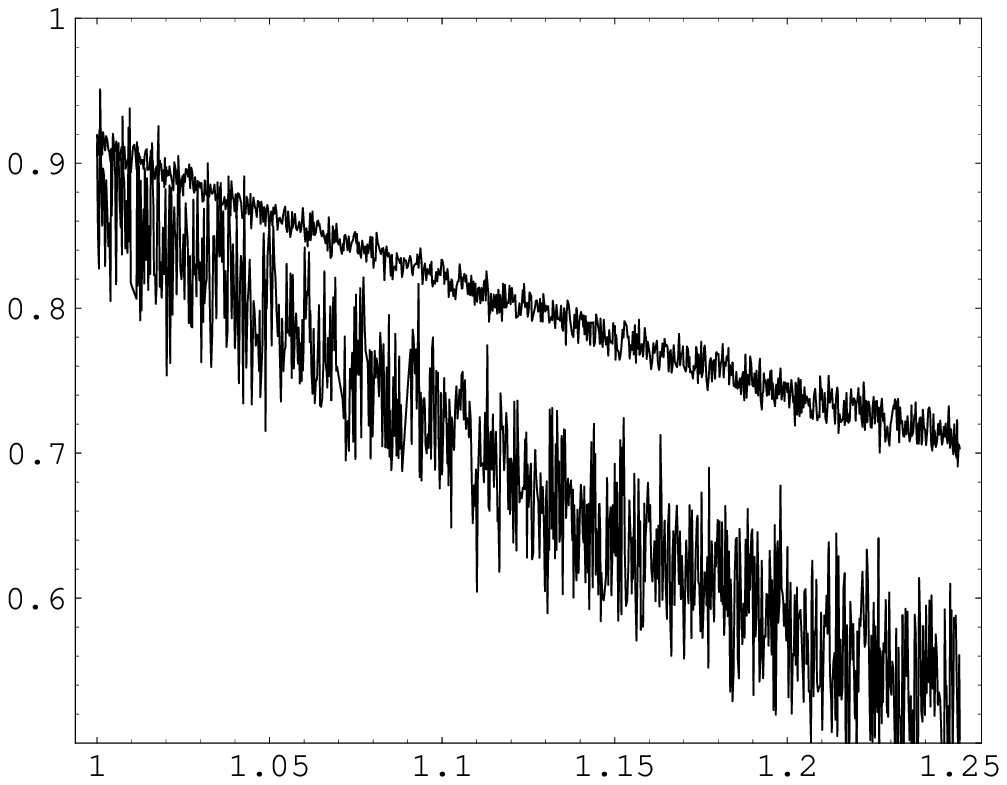}}
\end{center}
\vspace*{-16pt}
\fcaption{\footnotesize Upper and lower bounds on $(\omega+1)^2 \left|
h_{A_1}(\omega) \right|^2/(4\omega)$ in the regime $\omega \in
(1,1.25)$.  The curves include $O(1/m_Q^2)$ and perturbative
corrections.  HQET parameters are varying: ${\bar \Lambda} \in
(0.3,0.5)\; {\rm GeV}$, $\lambda_1 \in (-0.1,-0.3)\; {\rm GeV^2}$,
$\rho_{1,2}={\cal T}_{1,2,3,4} \in (-0.125,0.125)\; {\rm GeV^3}$, and
$\lambda_2=0.12\; {\rm GeV^2}$.  Perturbative parameters are:
$m_b=4.8\; {\rm GeV}$, $m_c=1.4 {\rm GeV}$ and $\Delta \in (1,2)\;
{\rm GeV}$.}
\vspace*{-16pt}
\end{figure}
The bounds at zero recoil are roughly $0.84 \lsim f(1) \lsim 0.94$
where $O(\as^2 \beta_0)$ corrections at zero recoil are also included.
Here the uncertainty is largely due to poor knowledge in the HQET
parameters.  The upper bounds solely depend upon $\ld_1$ and ${\bar
\Ld}$, while the determination of the lower bounds are also affected
by parameters showing up at $O(1/m_Q^3)$.  Also, in order to have a
better understanding of the bounds at the order being considered and
at large $\omega$, one should include $O(\as^2\beta_0)$ corrections to
the full spectrum.

{\it Acknowledgments}: The author is grateful to Adam Leibovich for the
collaboration and to Pauline Gagnon and Zoltan Ligeti for their kind
arrangement of the talk.  He also thanks Mike Luke and Matthias
Neubert for interesting discussions on the subject.  This research is
supported by the Department of Energy under Grant
No. DE-FG02-91ER40682.

\end{document}